\documentclass[reprint,superscriptaddress,showpacs,amsmath,amssymb,aps,prb,]{revtex4-1}

\usepackage{graphicx}
\usepackage{dcolumn}
\usepackage{bm}
\usepackage{natbib}

\newcommand{\av}[1]{\left\langle #1 \right\rangle} 
\newcommand{\tr}[1]{{\rm{Tr}}\left\{ #1\right\}}
\newcommand{\mc}[1]{\mathcal #1}

\newcommand{\bs}[1]{\boldsymbol{#1}}

\newlength{\graphwid}
\def\showgraph#1#2{
\settowidth{\graphwid}{\includegraphics[#1,clip=true]{#2}}
\parbox[c]{\graphwid}{\includegraphics[#1,clip=true]{#2}}}

\begin{document}

\title{Nonequilibrium Spin Noise and Noise of Susceptibility}

\author{P. Schad}

\affiliation{Institut f\"ur Theorie der Kondensierten Materie,
  Karlsruhe Institute of Technology, D-76131 Karlsruhe, Germany}

\author{B.N. Narozhny} 
\affiliation{Institut f\"ur Theorie der Kondensierten Materie,
  Karlsruhe Institute of Technology, D-76131 Karlsruhe, Germany}
\affiliation{National Research Nuclear University MEPhI (Moscow Engineering Physics Institute), 
  Kashirskoe shosse 31, 115409 Moscow, Russia }

\author{Gerd Sch\"on}
\affiliation{Institut f\"ur Theoretische Festk\"orperphysik, Karlsruhe
  Institute of Technology, D-76131 Karlsruhe, Germany}
\affiliation{Institut f\"ur Nanotechnologie, Karlsruhe Institute of
  Technology, D-76021 Karlsruhe, Germany}
  
  \author{A. Shnirman}
\affiliation{Institut f\"ur Theorie der Kondensierten Materie,
  Karlsruhe Institute of Technology, D-76131 Karlsruhe, Germany}

\date{\today}

\begin{abstract}
 We analyze out-of-equilibrium fluctuations in a driven spin system
 and relate them to the noise of spin susceptibility. In the spirit of
 the linear response theory we further relate the noise of
 susceptibility to a $4$-spin correlation function in equilibrium. We
 show that, in contrast to the second noise (noise of noise), the
 noise of susceptibility is a direct measure of non-Gaussian
 fluctuations in the system. We develop a general framework for
 calculating the noise of susceptibility using the Majorana
 representation of spin-1/2 operators. We illustrate our approach by a
 simple example of non-interacting spins coupled to a dissipative
 (Ohmic) bath.
\end{abstract}


\maketitle

Noise in electronic circuits provides information about the
microscopic structure of the system complementary to that obtained
from linear response transport
measurements~\cite{kogan,Blanter20001}. For electronic circuits, the
standard Johnson-Nyquist noise is intimately related to dissipative
processes with typical time scales of the order of picoseconds. At low
frequencies it is ``white'', i.e. frequency-independent. In contrast,
the ubiquitous $1/f$ noise is related to slow processes, e.g., to slow
rearrangements of impurities or the internal dynamics of two-level
systems~\cite{FengLeeStone}. Its power spectrum is commonly described
by the Hooge's law~\cite{Hooge69,Hooge76}, $S_V(f)\propto\overline
V^2/f$, where $\overline V$ is the average observed voltage. This
suggests that this noise could only be observed out of
equilibrium. But this was shown not to be the case by Voss and
Clarke~\cite{VossClarkePRL,VossClarkePRB}, who measured the
low-frequency fluctuations of the mean-square Johnson voltage in
equilibrium (i.e., the second noise or noise of noise) and showed that
these fluctuations possess a $1/f$-like spectrum.

Motivated by these experiments, Beck and Spruit~\cite{beckspruit}
calculated the variance of the Johnson-Nyquist noise and showed that
it comprised two contributions. The first one could be interpreted as
arising from resistance fluctuations with a $1/f$ spectrum. The
second, with a white spectrum, is intrinsic to any Gaussian
fluctuating quantity. Consequently the equilibrium $1/f$ noise could
only be observed at very low frequencies.

From a technical point of view the variance of noise is described by a
four-point correlation function~\cite{kogan,beckspruit}. Such objects
appear also in other physical contexts. For example,
Weissman~\cite{weissman88,weissman93} has proposed to distinguish the
droplet and hierarchical models of spin glasses by the properties of
the second noise, which can be expressed in terms of a particular
four-spin correlation function.

More recently, the problem of $1/f$ noise has attracted much attention
in the field of superconducting quantum devices. Flux noise
measurements initially performed with relatively large SQUIDs showed
the $1/f$ behavior~\cite{WellstoodClarke}. In the last decade similar
effect has been observed in nanoscale quantum
circuits~\cite{yoshihara06,kakuyanagi,bialczak07,McDermott,lanting,Yoshihara10,Wellstood11,Gustavsson11,Drung11,Anton12,Sank12,Lanting2014}. Remarkably,
the noise magnitude appears to be ``universal", i.e., of the same
order of magnitude for a wide range of device sizes. This noise is
believed to originate from an assemblage of spins localized at the
surface or interface
layers~\cite{KochClarke,FaoroIoffe,ChoiLee,Faoro12}. Indeed evidence
for the existence of surface spins was obtained in several dedicated
experiments~\cite{SendelbachMcDermott1,BluhmMoler}. All these
experimental findings are consistent with a surface spin density of
$\sigma_S\sim 5 \cdot 10^{17}\ \rm{m}^{-2}$.

Sendelbach \textit{et al.}\cite{SendelbachMcDermott2} recently
developed an alternative method to measure properties of flux noise,
namely by observing fluctuations in the inductance of SQUIDs. As the
spin contribution to the inductance is determined by the spin
susceptibility~\cite{McDermott} this experiment essentially amounts to
measuring the noise of the susceptibility. Technically this quantity
also corresponds to a four-spin correlation function, which however is
distinct from the correlation function desribing the second noise. To
the best of our knowledge, there is no consensus in the literature on
how to define noise of susceptibility. Some authors employ the
fluctuation-dissipation theorem and, thus, relate the noise of noise
and the noise of susceptibility~\cite{ChenYu,De}. This relation seems
to be justified in cases where the system is controlled by a slowly
fluctuating parameter, always remaining in a
quasi-equilibrium. Nevertheless, there is a need for a more general
definition of the noise of susceptibility at the microscopic level.

In this paper we focus on spin systems where we pursue the following
issues: (i) we give a general definition of noise of susceptibility in
terms of four-spin correlation functions and emphasize its distinction
from the second noise; (ii) as an illustration, we compute the noise
of susceptibility in a simple model of a single spin $1/2$ immersed in
a dissipative environment; (iii) in order to perform the above
calculation, we further develop a powerful
technique~\cite{Mao03,Shnirman03,sachdev,spencerd,spencer} based on
the Majorana-fermion representation of spin-$1/2$
systems~\cite{Martin,Tsvelik,Berezin77,shastri}; (iv) we use the above
results to estimate the noise of susceptibility for the model of
independent spins with a distribution of the relaxation rates that
widely used to describe $1/f$-noise (the Dutta-Horn
model~\cite{DuttaHorn}). The results of the latter calculation are
incompatible with the experimental results of
Ref.~\onlinecite{SendelbachMcDermott2} and we conclude that the
observed surface spins cannot be described by non-interacting models.

We find that the four-spin correlation function corresponding to the
noise of susceptibility vanishes if evaluated for Gaussian fluctuating
quantities. Hence a system of harmonic oscillators (photons or
phonons) would show no fluctuations of susceptibility. Therefore the
noise of susceptibility constitutes a direct measure of non-Gaussian
fluctuations. In contrast, the second noise is present in Gaussian
systems, where it is independent of frequency. Furthermore, in
non-Gaussian systems the second noise always contains this trivial
Gaussian contribution, which often masks the interesting non-Gaussian
noise making the latter very hard to
observe~\cite{Nguyen01,ClarkeUnpublished}.

As an illustration of our general arguments, we consider a model of
independent spins. The spin degrees of freedom are intrinsically
non-Gaussian even in the absence of spin-spin interactions. Indeed, in
this model we find a non-vanishing white noise of susceptibility, that
scales as $N/T^2$, where $N$ is the number of spins and $T$ the
temperature. At the same time the average susceptibility scales as
$N/T$, i.e., the fluctuations of susceptibility are small, as expected
for a non-interacting system.

\section{Qualitative arguments}

Let us illustrate the commonly used statistical concepts devoted to
noise by a simple example. Consider a random quantity $x$ with
probability distribution $P(x)=Z^{-1} \exp[-U(x)]$. It provides
complete information about $x$, which alternatively can be expressed
by specifying all moments $\langle x^n \rangle$ or all cumulants
$\langle\langle x^n\rangle\rangle$. For a symmetric distribution,
$P(x) = P(-x)$ such that $\langle x\rangle =0$, the noise of $x$ is
defined as~\cite{kogan}
\begin{equation}
\label{S1}
S_1 \equiv 2 \left\langle x^2\right\rangle.
\end{equation}
Up to the factor $2$, the noise is equal to the second cumulant of $x$
(which is here also equal to the second moment since $\langle x\rangle
=0$).

The second noise (or noise of noise) of $x$ is defined as~\cite{kogan}
\begin{equation}
\label{S2}
S_2 \equiv 2 \left(\left\langle x^4 \right\rangle - \left\langle x^2\right\rangle^2\right), 
\end{equation}
which is neither a higher moment nor a cumulant. This particular definition
is motivated by the measurement protocol~\cite{kogan}, in which the
time fluctuations of $x^2$ are recorded (see also Appendix~\ref{nn}).
For a Gaussian random quantity ($U$ is quadratic) one finds
\begin{equation}
\left\langle\left\langle x^4 \right\rangle\right\rangle = 0,\qquad  
\left\langle x^4 \right\rangle = 3  \left\langle x^2\right\rangle^2,\qquad 
S_2 = 4  \left\langle x^2\right\rangle^2.
\end{equation}

Now we perturb our system by a weak external field $B$, so that the
new distribution function reads $P_B(x)=Z_B^{-1}\exp[-U(x)+Bx]$. Then
the random quantity $x$ acquires a nonzero average value
\begin{equation}
\label{xb}
\left\langle x\right\rangle_B =Z_B^{-1}\frac{\partial Z_B}{\partial B} = \chi B 
+ {\cal O}(B^3), 
\end{equation}
where $\chi=\langle x^2\rangle$ is the corresponding linear
susceptibility. The second moment of $x$ acquires an additional field
dependent term, i.e.,
\begin{equation}
\label{x2b}
\left\langle x^2 \right\rangle_B =Z_B^{-1}\frac{\partial^2 Z_B}{\partial B^2} 
= \left\langle x^2 \right\rangle+ (\chi^2  + a) B^2 + {\cal O}(B^4),
\end{equation}
where
\begin{equation}
\label{a}
a = \frac{1}{2} \left(\left\langle x^4 \right\rangle - 
3\left\langle x^2 \right\rangle^2\right).
\end{equation}
For a Gaussian distribution one has $a=0$. Thus, the quantity $a$
given by Eq.~(\ref{a}) is a measure of the non-Gaussian nature of
fluctuations. At the same time, $a$ is proportional to the 4-th
cumulant of $x$ and, therefore, is inequivalent to the second noise
$S_2$.

In typical
experiments~\cite{VossClarkePRL,VossClarkePRB,SendelbachMcDermott2},
the fluctuating quantity is time-dependent and instead of the averages
(\ref{x2b}) one has to consider correlation functions (see below).
The noise is characterized by the spectral
power~\cite{kogan,Blanter20001} which is the Fourier transform of the
corresponding correlation function evaluated in the presence of the
external field.  Such an analysis of the experimental data is usually
performed over a reasonably long, but necessarily limited time
interval. Repeating the analysis over a large number of such intervals
one may find that the susceptibility $\chi$ itself takes different
values at different times~\cite{SendelbachMcDermott2}. [Note, that
  this averaging is no longer described by the above model
  distribution $P(x)$. Within this simple model the susceptibility
  defined in Eq.~(\ref{xb}) does not fluctuate.] Averaging over the
fluctuating values of the susceptibility one finds its mean value. It
is then tempting to use this averaged susceptibility in
Eq.~(\ref{x2b}) and interpret the quantity $a$ as the noise of the
susceptibility~\cite{SendelbachMcDermott2}. At this point one has to
be careful, as there is no guarantee that $a$ is positive. In fact, it
is well-known in the theory of shot noise~\cite{Blanter20001,butt,les}
that out of equilibrium the noise may be lower than the equilibrium
noise at the same temperature.

As an illustration for such a negative non-equilibrium contribution to
the noise we consider a single spin $1/2$ subject to an external
magnetic field. If one is interested in equal-time correlators one can
use the above arguments with $x$ replaced by $\hat{S}_z$. Now, the
square of the spin operator is simply proportional to the identity
operator independent of whether the field is applied or
not. Consequently, $\langle \hat{S}_z^2\rangle_B=\langle
\hat{S}_z^2\rangle=1/4$ and $a=-\chi^2=-1/16$.

Thus one might expect the non-equilibrium spin fluctuations to be
described by the negative quantity (\ref{a}) which appears to be
inconsistent with its interpretation as noise of the
susceptibility. In what follows, we provide a proper microscopic
definition of the noise of susceptibility corresponding to the
experimental protocol proposed in
Ref~\onlinecite{SendelbachMcDermott2}.

\section{Non-equilibrium spin fluctuations and noise of the susceptibility}
\label{sec:suscor}

We now generalize the above arguments to the case of a quantum spin
system. Assume a coupling $\hat{H}_I = - \hat{\bs{S}} \bs{B}(t)$,
where $\hat{\bs{S}}$ is the spin operator and $\bs{B}$ is a magnetic
field. A traditional way of describing the response of the system to a
weak external perturbation is the spin susceptibility, which relates
the applied field to the resulting magnetization,
$M_i\equiv\langle\hat{S}_i(t)\rangle=\int dt'\chi_{ij}(t,t')B_j(t')$,
with the susceptibility given by the Kubo formula~\cite{Mahan}
\begin{equation}
\label{kubo}
\chi_{ij}(t, t') = i \theta(t-t') 
\left\langle \left[ \hat S_i(t), \hat S_j(t') \right]\right\rangle.
\end{equation}
Here the spin operators must be in the Heisenberg representation with
respect to the Hamiltonian of the system in the absence of the
field. In isotropic systems, the susceptibility tensor is diagonal
$\chi_{ij} = \chi \delta_{ij}$.

Time-dependent magnetization fluctuations can then be described by
a power spectrum (or spectral density) that in the simplest case can
be related to the imaginary part of the susceptibility (\ref{kubo})
with the help of the fluctuation-dissipation theorem~\cite{kogan}
\begin{eqnarray}
\label{noise}
&&
S_{M}(\omega) \equiv \langle \hat S_z(t)\hat S_z(t') +
\hat S_z(t')\hat S_z(t) \rangle_\omega
\\
&&
\nonumber\\
&&
\qquad\quad \;
= 2\coth\frac{\omega}{2T} {\rm Im} \chi(\omega).
\nonumber
\end{eqnarray}
The quantity (\ref{noise}) is a generalization of Eq.~(\ref{S1}). 

Systematic calculations are often facilitated by using
field-theoretical techniques. Real-time fluctuations, especially in
the presence of an external field, can be conveniently described
within the Keldysh formalism~\cite{Kamenev}. In this formalism, the
spin susceptibility (\ref{kubo}) has the form
\begin{eqnarray}
\label{chikel}
&&
\chi_{ij}(t,t') = i \av{{\cal T}_K \hat S_i^{cl}(t) \hat S_j^q(t')}_0
\\
&&
\nonumber\\
&&
\qquad\qquad
= -i \left.
\frac{\delta^2{\cal Z}[\lambda^{cl},\lambda^q]}
{\delta\lambda_i^q(t)\delta\lambda_j^{cl}(t')}\right|_{\lambda=0},
\nonumber
\end{eqnarray}
where ${\cal T}_K$ denotes time ordering along the Keldysh contour,
and ${\cal Z}[\lambda^{cl},\lambda^q]$ is the Keldysh partition
function with the source terms $\lambda^{cl(q)}$ included. The
subscripts $q$ and $cl$ on both the spin operators and source fields
refer to the so-called ``quantum'' and ``classical''
variables~\cite{Kamenev}. They are related to the fields belonging to
the upper (u) and lower (d) branch of the Keldysh contour according to
\begin{eqnarray*}
&&
\hat S_i^{cl}= \frac{1}{\sqrt{2}}\left( \hat S_i^u + \hat S_i^d\right),\quad 
\hat S_i^q= \frac{1}{\sqrt{2}}\left(\hat S_i^u - \hat S_i^d\right) 
\\
&&
\\
&&
\lambda_i^{cl}= \frac{1}{\sqrt{2}} \left( \lambda_i^u + \lambda_i^d\right), 
\quad \lambda_i^q= \frac{1}{\sqrt{2}} \left(\lambda_i^u - \lambda_i^d\right).
\end{eqnarray*}
The ``classical'' source term defined in this way describes the
physical probing field, $\lambda_i^{cl} \equiv \sqrt{2} B_i$, while
the ``quantum'' term is only needed to construct the correlation
function and is set to zero at the end of the calculation. Once the
susceptibility is obtained, we can use Eq.~(\ref{noise}) to find
the noise spectrum.

Alternatively, we can characterize fluctuations of the magnetization
by directly evaluating the second moment of the spin in the presence
of the perturbation, generalizing Eq.~(\ref{x2b}). Without loss of
generality, we can assume that the external field is applied along $z$
direction. The symmetrized second moment of the $z$-component of the
physical spin is then given by
\begin{eqnarray}
\label{Scor}
&&
\av{\hat S_z(t_1) \hat S_z(t_2)+ \hat S_z(t_2) \hat S_z(t_1)}_B =
\\ 
&&
\nonumber\\
&&
\qquad
=\av{{\cal T}_K\hat S_z^{cl}(t_1) \hat S_z^{cl} (t_2)}_B =
- \left.
\frac{\delta^2Z[\lambda_z^q,B]}{\delta\lambda_z^q(t_1)\delta\lambda_z^q(t_2)}\right|_{\lambda_z^q=0} 
\nonumber\\ 
&&
\nonumber\\
&&
\qquad\qquad
 = \av{{\cal T}_K\hat{S}_z^{cl}(t_1) \hat{S}_z^{cl}(t_2) e^{ i\int dt'\sqrt{2}B\hat{S}_z^q } }.
\nonumber
\end{eqnarray}
Note, that for a spin $1/2$ the moment (\ref{Scor}) at equal times
$t_1=t_2$ is given by a $B$-independent constant (which is equal to
$1/2$).

For weak external fields, we may expand the quantity (\ref{Scor}) in
a power series in $B$,
\begin{eqnarray}\label{SzSzB}
&&
\qquad
\av{\hat {\cal T}_K S_z^{cl}(t_1) \hat S_z^{cl} (t_2)}_B  =  S_{M}(t_1-t_2) 
\\
&&
\nonumber\\
&&
\qquad\quad
+  \int dt_1'dt_2' C_{\chi}(t_1,t_1',t_2,t_2') B(t_1') B(t_2') 
+ {\mc O}\left(B^4\right).
\nonumber
\end{eqnarray}
The first term in Eq.~(\ref{SzSzB}) corresponds to the equilibrium
noise (\ref{noise}) as it should be: the noise is characterized by the
second cumulant of the fluctuating quantity in the absence of the
applied field, similar to Eqs.~(\ref{S1}) and (\ref{noise}). 
For the spin $1/2$, it obeys the ``sum rule'' $S_M(0)=1/2$.

The second term in Eq.~(\ref{SzSzB}) contains the four-point
correlation function
\begin{eqnarray} 
\label{SCF}
&&
C_\chi(t_1,t_1',t_2,t_2') = -\left.\frac{\delta^4Z[\lambda^{cl},\lambda^q] }
{\delta\lambda_z^q(t_1)\delta\lambda_z^{cl}(t_1')\delta\lambda_z^q(t_2)\delta\lambda_z^{cl}(t_2')}
\right|_{\lambda=0} 
\nonumber\\
&&
\nonumber\\
&&
\qquad\qquad
= - \av{ {\mc T}_K \hat S_z^{cl}(t_1) \hat S_z^{q}(t_1') \hat S_z^{cl}(t_2) \hat S_z^{q}(t_2')} ,
\end{eqnarray}
which one can split into Gaussian and non-Gaussian parts 
\[
C_\chi =C_\chi^G + C_\chi^{NG}. 
\]
The Gaussian part is readily obtained by a pair-wise averaging of the
spin operators:
\begin{eqnarray} 
\label{Gdecomp}
&&
C^G_\chi(t_1,t_1',t_2,t_2')=
\\
&&
\nonumber\\
&&
\qquad\qquad
=-
\av{{\mc T}_K\hat S_z^{cl}(t_1)\hat S_z^{q}(t_1')}
\av{{\mc T}_K\hat S_z^{cl}(t_2)\hat S_z^{q}(t_2')}
\nonumber \\
&&
\nonumber \\
&&
\qquad\qquad\quad
-
\av{ {\mc T}_K \hat S_z^{cl}(t_1) \hat S_z^{q}(t_2')}
\av{ {\mc T}_K  \hat S_z^{cl}(t_2) \hat S_z^{q}(t_1') }
\nonumber \\
&&
\nonumber \\
&&
\qquad\quad
= \chi_{zz}(t_1,t_1') \chi_{zz}(t_2,t_2') + \chi_{zz}(t_1,t_2') \chi_{zz}(t_2,t_1') .
\nonumber
\end{eqnarray}
Note the absence of a contribution involving two ``quantum''
fields. Such terms vanish since the correlator of the two ``quantum''
fields is always zero.

As mentioned above, for the spin $1/2$ the moment (\ref{Scor}) at
equal times $t_1=t_2=t$ is equal to $1/2$ independently of the
magnetic field and therefore $C_\chi(t,t_1',t,t_2')=0$. Clearly, the
Gaussian contribution (\ref{Gdecomp}) does not satisfy this ``sum
rule'' and thus there must be a non-Gaussian contribution
$C_\chi^{NG}$ as well. More generally, the Wick's theorem does not
hold for spin operators, reflecting the fact that their algebra is
non-Abelian. This quantity cannot be expressed in terms of the
averaged susceptibilities and has to be evaluated specifically for
each system.

In a typical experiment~\cite{kogan,SendelbachMcDermott2}, the system
is probed with a harmonic perturbation,
\[
B(t) = B_0\cos(\omega_0 t).  
\]
The susceptibility is then measured using lock-in techniques, which
amounts to obtaining the average of the following operator
\begin{equation}
\label{chin}
\hat{\chi}_\varphi(\tau_n |\omega_0,\Delta\omega) =\frac{1}{B_0 T_\chi} 
\int\limits_{\tau_n-\frac{T_\chi}{2}}^{\tau_n+\frac{T_\chi}{2}}dt 
\cos(\omega_0 t - \varphi)\,\hat S_{z,B}(t).
\end{equation} 
The measurement is performed for a time period $T_\chi$ centered
around $\tau_n$. This defines the measurement bandwidth
$\Delta\omega\equiv 2\pi/T_\chi\ll\omega_0$, chosen to be much smaller
that $\omega_0$. The phase $\varphi$ allows discriminating between the
in-phase ($\varphi=0$) and the out-of-phase ($\varphi=\pi/2$)
response, corresponding to the real and imaginary parts of the
susceptibility, respectively. In practice, in every time bin one finds
a different result and the average susceptibility is obtained by
averaging over the time bins.

Treating the result of susceptibility measurements in each time bin as
a fluctuating quantity (as it is in the
experiment~\cite{SendelbachMcDermott2}), one can define its second
moment or noise of susceptibility as follows
\begin{eqnarray}
\label{c2}
&&
\chi^{(2)}_{\varphi_1,\varphi_2}(\tau_1,\tau_2|\omega_0,\Delta\omega) = 
\\
&&
\nonumber\\
&&
\qquad\qquad
=
\av{\chi_{\varphi_1}(\tau_1)\chi_{\varphi_2}(\tau_2) + 
\chi_{\varphi_2}(\tau_2)\chi_{\varphi_1}(\tau_1)} 
\nonumber\\
&&
\nonumber\\
&&
\qquad\qquad\qquad\qquad
- 2\av{\chi_{\varphi_1}(\tau_1)}\av{\chi_{\varphi_2}(\tau_2)}.
\nonumber
\end{eqnarray}
Using the explicit form (\ref{chin}), we find
\begin{eqnarray} 
\label{biSCF}
&&
\chi^{(2)}_{\varphi_1,\varphi_2}(\tau_1,\tau_2|\omega_0,\Delta\omega)=
\frac{1}{B_0^2 T_\chi^2}
\int\limits_{\tau_1-\frac{T_\chi}{2}}^{\tau_1+\frac{T_\chi}{2}}dt_1
\int\limits_{\tau_2-\frac{T_\chi}{2}}^{\tau_2+\frac{T_\chi}{2}}dt_2
\nonumber\\
&& 
\nonumber\\
&&
\qquad\qquad\quad
\times
\cos(\omega_0 t_1 - \varphi_1) \cos(\omega_0 t_2 - \varphi_2) 
\nonumber\\ 
&&
\nonumber\\
&&
\qquad\qquad\quad
\times \av{\av{\hat{S}(t_1) \hat{S}(t_2)+\hat{S}(t_2) \hat{S}(t_1)}}_{B} .
\end{eqnarray}
In contrast to Eq.~(\ref{chin}), the averaging in Eq.~(\ref{biSCF})
has been already performed (as we are not interested in higher
moments). The double angle brackets in Eq.~(\ref{biSCF}) indicate the
2-nd cumulant, which is obtained by subtracting two times the product
of averages, i.e., $2\av{\hat{S}(t_1)}\av{\hat{S}(t_2)}$, as in
Eq.~(\ref{c2}).

We now use the expansion (\ref{SzSzB}) for the symmetrized average
(\ref{Scor}) and decompose the second moment of susceptibility
Eq.~(\ref{biSCF}) into two parts
\[
\chi^{(2)} = \chi^{(2)}_{eq} + \chi^{(2)}_{ne}.
\]
The first term $\chi^{(2)}_{eq}$ describes the equilibrium
magnetization noise $S_{M}$ in the absence of the external field:
\begin{eqnarray} 
\label{biSCFeq}
&&
\chi^{(2)}_{eq,\varphi_1,\varphi_2}(\tau_1,\tau_2|\omega_0,\Delta\omega)=
\frac{1}{B_0^2 T_\chi^2}
\int\limits_{\tau_1-\frac{T_\chi}{2}}^{\tau_1+\frac{T_\chi}{2}}dt_1
\int\limits_{\tau_2-\frac{T_\chi}{2}}^{\tau_2+\frac{T_\chi}{2}}dt_2
\nonumber\\
&& 
\nonumber\\
&&
\qquad
\times
\cos(\omega_0 t_1 - \varphi_1) \cos(\omega_0 t_2 - \varphi_2) 
S_{M}(t_1-t_2).
\end{eqnarray}
The corresponding noise spectrum is given by the Fourier transform of
$\chi^{(2)}_{eq}$
\begin{eqnarray} 
\label{ChiEq}
&&
\chi^{(2)}_{eq,\varphi_1,\varphi_2}(\nu|\omega_0,\Delta\omega) = 
\frac{1}{4B_0^2} {f}\left(\frac{\pi\nu}{\Delta\omega}\right) 
\\
&&
\nonumber\\
&&
\qquad 
\times 
\left\{\cos(\varphi_1-\varphi_2)\left[S_{M}(\omega_0+\nu)+S_{M}(\omega_0-\nu)\right]\right.
\nonumber\\
&&
\nonumber\\
&&
\qquad\quad\quad
\left.
-i\sin(\varphi_1-\varphi_2)\left[S_{M}(\omega_0+\nu)-S_{M}(\omega_0-\nu)\right] \right\} 
\nonumber\\
&&
\nonumber\\
&&
\qquad\qquad\qquad\qquad
+{\cal O}\left(\frac{\Delta\omega}{\omega_0}\right),
\nonumber
\end{eqnarray}
where $f(x)\equiv\sin^2(x)/x^2$ restricts the frequency $\nu$ to be
small, ${\nu\lesssim\Delta\omega\ll\omega_0}$. The appearance of the
imaginary part in the noise spectrum (\ref{ChiEq}) reflects the fact
that cross-correlations between the real and imaginary parts of the
susceptibility do not possess any symmetry as functions of
time. Indeed, according to the definition (\ref{c2}) the noise of
susceptibility obeys the symmetry
\begin{equation}
\label{ccsym}
\chi^{(2)}_{\varphi_1,\varphi_2}(\tau_1,\tau_2) = 
\chi^{(2)}_{\varphi_2,\varphi_1}(\tau_2,\tau_1)
\end{equation}
and is a symmetric function of the two times $\tau_i$ only if
$\varphi_1=\varphi_2$. As a result, the noise of the real (or
imaginary) part of susceptibility is characterized by a real
spectrum, while the Fourier transform of the cross-correlator may
contain an imaginary part.

Turning to the non-equilibrium contribution $\chi^{(2)}_{ne}$,
composed of the second term of the expansion \eqref{SzSzB} substituted
into Eq.~\eqref{biSCF}, we note that only the non-Gaussian part
$C^{NG}_\chi$ of the correlation function (\ref{SCF})
contributes. This is because the Gaussian part (\ref{Gdecomp})
corresponds exactly to the subtracted product of the averages, i.e.,
the last term in Eq.~\eqref{c2}. Thus we obtain
\begin{eqnarray} 
\label{biSCFmeas}
&&
\chi^{(2)}_{ne,\varphi_1,\varphi_2}(\tau_1,\tau_2|\omega_0,\Delta\omega)=
\frac{1}{T_\chi^2} 
\\ 
&& 
\nonumber\\ 
&& 
\quad
\times\int\limits_{\tau_1-\frac{T_\chi}{2}}^{\tau_1+\frac{T_\chi}{2}}d
t_1 \int\limits_{\tau_2-\frac{T_\chi}{2}}^{\tau_2+\frac{T_\chi}{2}} d
t_2 \cos(\omega_0 t_1 - \varphi_1) \cos(\omega_0 t_2 - \varphi_2)
\nonumber\\ 
&& 
\nonumber\\ 
&& 
\qquad 
\times \int d t'_1 d t'_2
C^{NG}_\chi(t_1,t'_1,t_2,t'_2) \cos(\omega_0 t'_1)\cos(\omega_0 t'_2).
\nonumber
\end{eqnarray}

The time integrals in Eq.~(\ref{biSCFmeas}) can be simplified with the
help of the Fourier transform defined as follows
\begin{eqnarray}
\label{FourierChi}
&&
C^{NG}_\chi(t_1,t'_1,t_2,t'_2)= \int\frac{d\nu d\omega_1 d\omega_2}{(2\pi)^3}
C^{NG}_\chi(\nu,\omega_1,\omega_2)
\\
&& 
\nonumber\\
&&
\qquad\qquad\qquad\qquad
\times
e^{-i\nu(t_1-t_2)} e^{-i\omega_1(t_1-t'_1)} e^{-i\omega_2(t_2-t'_2)}.
\nonumber
\end{eqnarray}
As stated above, in this paper we are only interested in low-frequency
noise~\cite{SendelbachMcDermott2}
$\nu\ll\Delta\omega\ll\omega_0$. Focusing on contributions that are
slow functions of $\tau_1 - \tau_2$, we retain only the lowest
harmonics and find
\begin{eqnarray} 
\label{biSCFmeas4}
&&
\chi^{(2)}_{ne,\varphi_1,\varphi_2}(\nu|\omega_0,\Delta\omega)=
\frac{1}{16} {f}\left(\frac{\pi\nu}{\Delta\omega}\right) 
\\
&&
\nonumber\\
&&
\quad\quad
\times\left[
\sum_{\epsilon_1,\epsilon_2=\pm 1}e^{-i\epsilon_1\varphi_1}e^{-i\epsilon_2\varphi_2} 
C^{NG}_\chi(\nu,\epsilon_1\omega_0,\epsilon_2\omega_0)\right. 
\nonumber\\
&&
\nonumber\\
&&
\qquad\quad
\left.
+\sum_{\epsilon=\pm 1} e^{i\epsilon(\varphi_1-\varphi_2)}
C^{NG}_\chi(\nu-2\epsilon\omega_0,\epsilon\omega_0,-\epsilon\omega_0)
\right].
\nonumber
\end{eqnarray}
Thus, the non-equilibrium contribution to the noise of the
susceptibility is a probe of non-Gaussian fluctuations in the system,
in contrast to the second noise~\cite{kogan}. Below, we will
illustrate our general considerations by calculating $C_\chi^{NG}$ for
the simplest model system, i.e. a single spin coupled to a dissipative
environment.

\section{Susceptibility noise of a single spin}

\subsection{The model}

Let us now illustrate our general arguments using a simple example of
a single spin-$1/2$ coupled to a dissipative bath in the presence of a
magnetic field. In this Section, we calculate the four-spin
correlation functions (\ref{biSCFeq}) and (\ref{biSCFmeas4}) leaving
the discussion of the results and their relation to experiments for
the subsequent Section.

We model the bath by an isotropic, bosonic vector degree of freedom
$\hat{\vec X}$ coupled to the spin operator via the minimal
Hamiltonian
\begin{equation}
\label{sx}
H = \hat{\vec s} \cdot \hat{\vec X}.
\end{equation}
The physical properties of the bath can be encoded in the bosonic
correlation function. Here we have also chosen to incorporate the
coupling constant into the definition of the bosonic correlator.
Within the frame of the Keldysh formalism, the bosonic correlation
function is defined as
\begin{equation}
\label{pK}
\hat\Pi_{\alpha\beta}^{ab}(t,t')=\delta_{\alpha\beta} \Pi^{ab}(t,t') =
\tr{\mc{T}_K \hat X^{\alpha,a}(t) \hat X^{\beta,b}(t')},
\end{equation}
where Latin indices span the $2\times 2$ Keldysh space $a,b=cl,q$ and
Greek indices refer to the spin components $\alpha,\beta=x,y,z$. 
The bath being Ohmic means that the following relation holds:
\begin{equation}
\label{pr}
\Pi^R(\omega)-\Pi^A(\omega) = \lambda\omega,
\end{equation}
where $\lambda \ll 1$ is the effective coupling constant, and $\Pi^{R/A}$ are
the retarded and the advanced components of (\ref{pK}).  We
assume the bath to be in thermal equilibrium such that the Keldysh
component of the correlation function (\ref{pK}) is given by the
standard expression
\begin{subequations}
\label{px}
\begin{equation}
\label{pk}
\Pi^K(\omega)=\coth\frac{\omega}{2T} \left[\Pi^R(\omega)-\Pi^A(\omega)\right].
\end{equation}
\end{subequations}
Note that the model (\ref{sx}) is similar to the Kondo
model~\cite{PaaskeWoelfle2,Mao03,Tsvelik} in the high temperature
regime, $T\gg T_K$, where the latter effectively describes a spin
coupled to an Ohmic bath (\ref{pr}) .

\subsection{Majorana representation}

Our goal is to calculate a $4$-spin correlation function. In any
standard fermionic representation of the spin
operators~\cite{Tsvelik,Mahan}, $N$-point spin correlators correspond
to $2N$-point fermionic correlators. In our case, we would have to
evaluate an $8$-point fermionic correlation function which is
generally not an easy task. Fortunately, we can substantially simplify
calculations by introducing the so-called Martin's Majorana-fermion
representation~\cite{Tsvelik,Berezin77,Martin,spencerd,spencer,shastri,sachdev,Mao03,Shnirman03}
\begin{equation}
\label{mf}
\hat s^\alpha=-(i/2)\epsilon_{\alpha\beta\gamma} \eta_\beta  \eta_\gamma,
\qquad
\eta_\alpha^\dagger=\eta_\alpha,
\end{equation}
or explicitly
\[
\hat s^x= -i\eta_y \eta_z, \quad \hat s^y= -i\eta_z \eta_x, \quad \hat s^z= -i\eta_x \eta_y.
\]
The Majorana fermions obey the Clifford algebra
\begin{equation}
\label{mfa}
\{\eta_\alpha,\eta_\beta\}=\delta_{\alpha\beta}, \qquad \eta_\alpha^2=1/2.
\end{equation}
This representation is convenient since it explicitly preserves the
spin-rotation symmetry and allows for a straightforward formulation of
the field theory, while perfectly reproducing the $SU(2)$ algebra of
the operators $\hat s^\alpha$.

At the same time, the above representation is not ``exact'' as the
Hilbert space of the Majorana triplet is
ill-defined~\cite{spencer,Tsvelik,shastri,sachdev,spencerd,Shnirman03,Mao03}.
In order to build the proper fermionic Hilbert space, one can increase
the number of Majorana fermions in the theory making it even. Adding
an additional Majorana fermion, we may build a four-dimensional
Hilbert space, twice as large as the original Hilbert space of the
spin. Thus the Majorana representation possesses extra states, not
present in the original model. This is a well-known problem
\cite{Tsvelik,shastri,sachdev} that can be resolved on the basis of
the known~\cite{spencerd} but not widely appreciated conjecture: {\it
  The fact that the Hilbert space is bigger that the 2-dimensional
  spin-1/2 Hilbert space has no effect on the spin correlation
  functions}. This can be understood as follows. The Majorana Hilbert
space can be roughly thought of as consisting of two copies of the
physical spin~\cite{Shnirman03}. Any operator of any physical quantity
operates only within a two-dimensional subspace corresponding to one
of the two spin copies. The remaining subspace does contribute to the
partition function, but this contribution is limited to a
multiplicative factor that cancels out from any correlation
function. A rigorous proof of this statement will be published
elsewhere~\cite{Annals}.

In the Majorana representation the Hamiltonian (\ref{sx}) takes the
form
\begin{equation} 
\label{Hint}
H= -(i/2)\epsilon_{\alpha\beta\gamma} \hat X^\alpha \eta_\beta  \eta_\gamma.
\end{equation}
Any spin correlation function can now be represented as a correlation
function of the Majorana fermions \cite{Shnirman03,Mao03,Annals}. For
example, the four-point function is given by
\begin{eqnarray*}
&&
\av{\hat s^{\alpha}(t_1)\hat s^{\beta}(t_1')\hat s^{\gamma}(t_2)\hat s^{\delta}(t_1')} =
\\ 
&&
\\
&&
\qquad\qquad\qquad\qquad
= (1/4)\av{ \eta_\alpha(t_1) \eta_\beta(t_1') \eta_\alpha(t_2) \eta_\beta(t_2')}.
\end{eqnarray*}
This relation demonstrates the strength of the Majorana representation
(\ref{mf}): the four-point spin correlator is given by the four-point
correlator of the Majorana fermions. However, in order to extend this
correspondence to the time ordered correlations functions on the
Keldysh contour, we need to take care of the time-ordering operator
${\cal T}_K$. In terms of spin operators, the time ordering is similar
to that of bosons [see, e.g. Eqs.~(\ref{chikel}) and (\ref{SCF})].
Yet, for the fermionic operators $\eta_\alpha$ the time ordering is
different. Therefore, it is convenient to multiply every Majorana
fermion in the above correlator by $-i m$, where $m$ is the fourth
Majorana fermion (needed anyway to construct the Hilbert space).
Given the algebra (\ref{mfa}), this operation does not change the
correlation function (which is effectively multiplied by $1/4$). With
respect to the time ordering the bilinear terms $-im\eta_\alpha$
behave as bosons similarly to the spin operators. The correlation
function (\ref{SCF}) then takes the form
\begin{eqnarray} 
\label{Cichi}
&&
C_\chi(t_1,t_1',t_2,t_2')= - 
\av{{\mc T}_K\hat s^{cl}(t_1)\hat s^{q}(t_1')\hat s^{cl}(t_2)\hat s^{q}(t_1')} 
\\
&&
\nonumber\\
&&
\qquad\qquad\quad
=-\av{ {\mc T}_K(\eta_x m)^{cl}_{t_1}(\eta_x m)^{q}_{t_1'}
(\eta_x m)^{cl}_{t_2}(\eta_x m)^{q}_{t_2'}}.
\nonumber
\end{eqnarray}

At this stage one might get an impression that we have achieved
nothing, as we are back to an $8$-fermion correlation
function. However, the operator $m$ commutes with the Hamiltonian of
the system. The Green's functions corresponding to $m$ remain
``bare'', which is a great simplification.

\subsection{Diagrammatic expansion}

\begin{figure}
\includegraphics[scale=1.1]{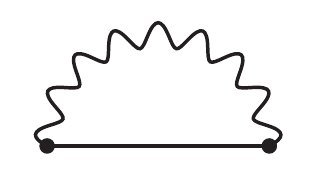}
\caption{The leading contribution to the self-energy. The wavy line
  refers to the bosonic correlator (\ref{px}) and the solid line to
  $G_{0,\alpha}$, see Eq.~(\ref{gd}).}
\label{SelfEnergy}
\end{figure}

Having defined the model in the Majorana representation we can now
proceed using the usual diagrammatic technique, which was not possible
for the original spin operators. The peculiarity of the single-spin
model is that the spin has no Hamiltonian in the absence of the bath
(and the magnetic field). Therefore, the ``free'' Green's functions of
the Majorana fermions are
\begin{eqnarray}
\label{gd}
&&
G_{0,\alpha}^R(t,t')=-i\av{{\mc T}_K\eta^{cl}_\alpha(t)\eta^q_\alpha(t')}=-i\Theta(t-t'), 
\\
&&
\nonumber\\
&&
D^R(t,t')= -i \av{ {\mc T}_K  m^{cl}(t) m^q(t')}= -i\Theta(t-t').
\nonumber
\end{eqnarray}
The coupling of the Majorana fermions $\eta_\alpha$ to the bath
(\ref{sx}) is then described by means of a ``self-energy'', which can
be obtained as a saddle-point solution in the path-integral
formulation of the theory \cite{Annals}. At high enough temperatures,
the leading contribution to the self-energy is graphically depicted in
Fig.~\ref{SelfEnergy} [where the wavy line refers to the bosonic
  correlator (\ref{px}) and the solid line to $G_{0,\alpha}$] and is
given by
\begin{equation}
\Sigma^R_\alpha= -i\Gamma = -i \lambda T.
\end{equation}
Consequently, the Green's functions of the Majorana fermions
$\eta_\alpha$ in the model (\ref{sx}) take the simple form
\begin{eqnarray}
\label{gra}
&&
G^{R/A}_\alpha(\omega)= \frac{1}{\omega \pm i\Gamma} 
\\
&&
\nonumber\\
&&
G^K_\alpha(\omega)= -\frac{2i\Gamma}{\omega^2+\Gamma^2}\tanh\frac{\omega}{2T}.
\nonumber
\end{eqnarray}

\begin{figure}
\showgraph{width=.2\textwidth}{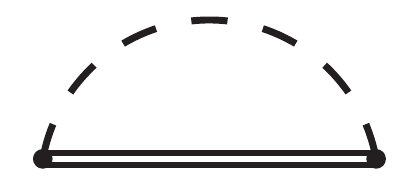}
\caption{The Majorana representation of the two-spin correlation
  function. This representation is exact as the additional Majorana
  fermion $m$ does not enter the Hamiltonian. The double solid line
  corresponds to the ``dressed'' Green's functions (\ref{gra}) of the
  Majorana fermions $\eta_\alpha$ and the dashed line refers to the
  Green's function $D$ of the non-interacting Majorana fermion $m$,
  see Eq.~(\ref{gd}).}
\label{eqd}
\end{figure}

The equilibrium noise of the susceptibility~(\ref{ChiEq}) is described
by the two-point function corresponding to the diagram in
Fig.~\ref{eqd}, where the double solid line corresponds to the
``dressed'' Green's functions (\ref{gra}) of the Majorana fermions
$\eta_\alpha$ and the dashed line refers to the Green's function $D$
of the non-interacting Majorana fermion $m$, see Eq.~(\ref{gd}). The
noise of the real part of susceptibility is identical with that of the
imaginary part
\begin{subequations}\label{ChiEqAll}
\begin{eqnarray}
\label{eqcontr}
&&
\chi^{(2)}_{eq,00}(\nu\vert \omega_0,\Delta\omega) =
\chi^{(2)}_{eq,\frac{\pi}{2},\frac{\pi}{2}}(\nu\vert \omega_0,\Delta\omega) =
\\
&&
\nonumber\\
&&
\quad
=
\frac{1}{4B_0^2} {f}\left(\frac{\pi\nu}{\Delta\omega}\right) 
\left[\frac{\Gamma}{\Gamma^2+(\omega_0+\nu)^2} +
\frac{\Gamma}{\Gamma^2+(\omega_0-\nu)^2}\right], 
\nonumber
\end{eqnarray}
and is purely real in accordance with the symmetry (\ref{ccsym}). In
contrast, the cross-correlations are characterized by the purely
imaginary noise spectrum
\begin{eqnarray}
\label{eqcross}
&&
\chi^{(2)}_{eq,0,\frac{\pi}{2}}(\nu\vert \omega_0,\Delta\omega) = 
\frac{i}{4B_0^2} {f}\left(\frac{\pi\nu}{\Delta\omega}\right) 
\\
&&
\nonumber\\
&&
\qquad\qquad\qquad\quad
\times
\left[\frac{\Gamma}{\Gamma^2+(\omega_0+\nu)^2} -
\frac{\Gamma}{\Gamma^2+(\omega_0-\nu)^2}\right].
\nonumber
\end{eqnarray}
\end{subequations}
The result (\ref{eqcross}) is an odd function of $\nu$, such that
integrating over the frequency would yield zero, corresponding to the
absence of cross-correlations between the real and imaginary parts of
susceptibility at equal times.

\begin{figure}
\showgraph{width=.2\textwidth}{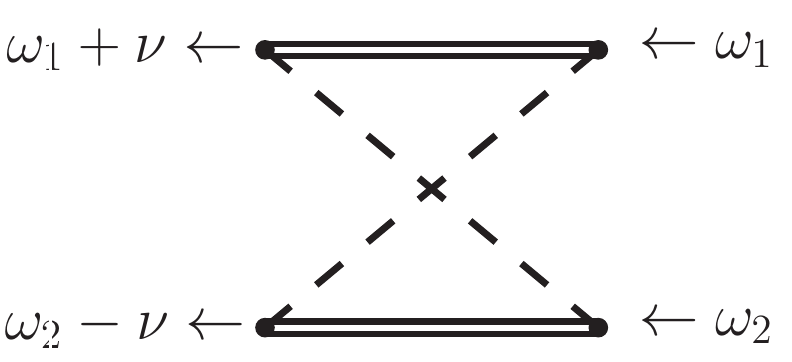} + \showgraph{width=.1\textwidth}{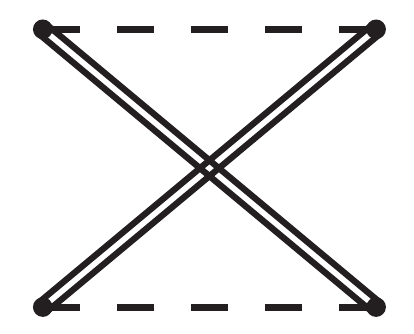} 
\\
+\showgraph{width=.1\textwidth}{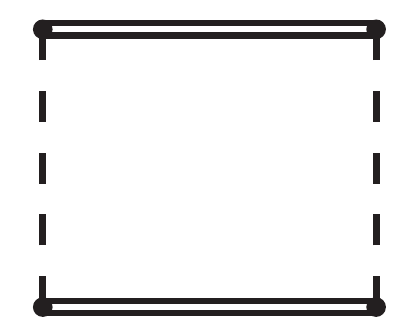} + \showgraph{width=.1\textwidth}{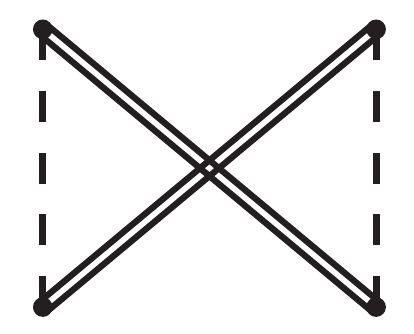} 
\\ 
+ \showgraph{width=.1\textwidth}{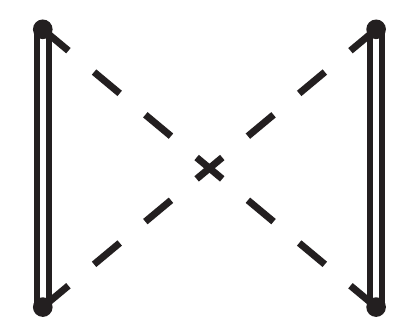} + \showgraph{width=.1\textwidth}{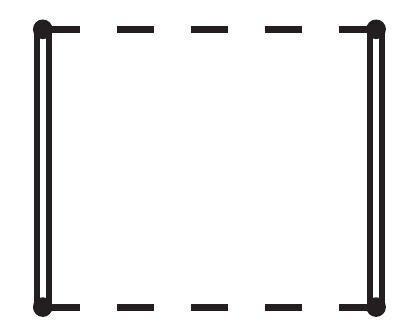} 
\caption{The leading contribution to non-equilibrium noise of susceptibility.} 
\label{lodiagrams}
\end{figure}

\begin{figure}
\showgraph{width=.15\textwidth}{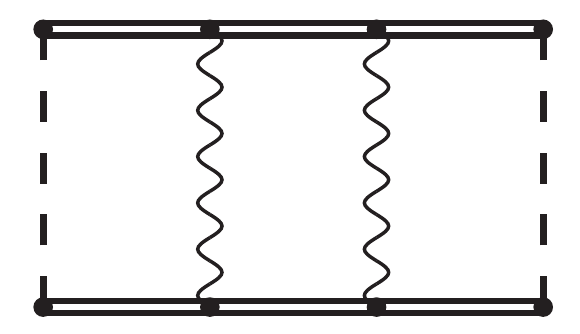} +
\showgraph{width=.15\textwidth}{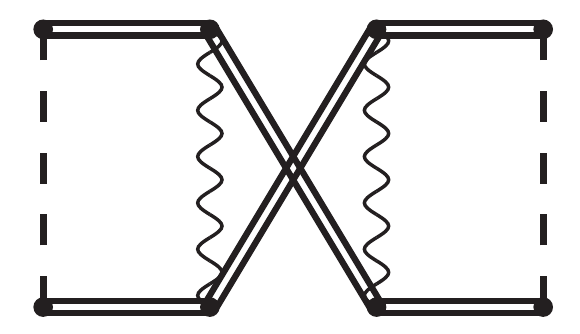} 
\caption{Examples of higher-order diagrams.}
\label{C1cor}
\end{figure}

Now we turn to the description of the more interesting,
non-equilibrium noise of susceptibility. For this purpose, we need to
evaluate the correlation function (\ref{Cichi}). Within the
path-integral formulation of the theory \cite{Annals}, we may
integrate out the bosonic bath and obtain the effective action for the
Majorana fermions. This action can be formally expanded around the
saddle-point solution. The result can be graphically presented in
diagrammatic form, see Figs.~\ref{lodiagrams} and \ref{C1cor}. The
external frequencies in these diagrams correspond to the Fourier
transform~(\ref{FourierChi}).

At high enough temperatures, $T\gg{T_K},\omega_1,\omega_2,\nu,\Gamma$,
the leading contribution to the correlation function (\ref{Cichi}) is
described by the diagrams in Fig.~\ref{lodiagrams} and is given by
\begin{eqnarray} 
\label{res1}
&& 
C^{NG}_\chi(\nu,\omega_1,\omega_2) = 
\\
&&
\nonumber\\
&&
= \frac{i\Gamma^2}{8T^2} \frac{\omega_1+\omega_2 + 2 i \Gamma}
{(\omega_1+i\Gamma)(\omega_2+i\Gamma)(\omega_1+\nu+i\Gamma)(\omega_2-\nu+i\Gamma)}.
\nonumber
\end{eqnarray}
Higher-order diagrams (see, e.g., Fig.~\ref{C1cor}) may be neglected
as long as the coupling constant $\lambda$ remains small. We have
ruled out the possibility that ladder diagrams might contribute to the
lowest order in $\lambda$. The details of this calculation will be
presented in a separate publication~\cite{Annals}.

Substituting the result (\ref{res1}) into Eq.~(\ref{biSCFmeas4}), we
find the non-equilibrium noise spectrum of the spin susceptibility in
the model (\ref{sx}). For the noise of the real part of the
susceptibility we find
\begin{subequations}\label{ChiNeqAll}
\begin{eqnarray}
\label{res2}
&&
\chi^{(2)}_{ne,00}(\nu\vert \omega_0,\Delta\omega) =
\frac{{f}\left(\frac{\pi\nu}{\Delta\omega}\right) }{32 T^2}
\\
&&
\nonumber\\
&&
\qquad\qquad
\times
\frac{\Gamma^3\left[\omega_0^2-3(\Gamma^2+\nu^2)\right]}
{(\Gamma^2+\omega_0^2)\left[(\Gamma^2+\nu^2)^2+2(\Gamma^2-\nu^2)\omega_0^2+\omega_0^4\right]}.
\nonumber
\end{eqnarray}
In contrast to the equilibrium contribution (\ref{eqcontr}), the noise
of the imaginary part of the susceptibility is inequivalent to
Eq.~(\ref{res2}) and is given by
\begin{eqnarray}
\label{res3}
&&
\chi^{(2)}_{ne,\frac{\pi}{2},\frac{\pi}{2}}(\nu\vert
\omega_0,\Delta\omega) = -
\frac{{f}\left(\frac{\pi\nu}{\Delta\omega}\right)}{32 T^2}
\\
&&
\nonumber\\
&&
\qquad\qquad
\times
\frac{\Gamma^3\left(\Gamma^2+5\omega_0^2+\nu^2\right)}
{(\Gamma^2+\omega_0^2)\left[(\Gamma^2+\nu^2)^2+2(\Gamma^2-\nu^2)\omega_0^2+\omega_0^4\right]}.
\nonumber
\end{eqnarray}
Finally, one can also compute the ``cross-correlation'' of the real
and imaginary parts of the susceptibility:
\begin{eqnarray}
\label{res4}
&&
\chi^{(2)}_{ne,0,\frac{\pi}{2}}(\nu\vert\omega_0,\Delta\omega) = -
\frac{{f}\left(\frac{\pi\nu}{\Delta\omega}\right)}{32 T^2}
\\
&&
\nonumber\\
&&
\qquad\qquad
\times
\frac{\omega_0\Gamma^2\left(3\Gamma^2-\omega_0^2+\nu^2 - 4 i \Gamma \nu \right)}
{(\Gamma^2+\omega_0^2)\left[(\Gamma^2+\nu^2)^2+2(\Gamma^2-\nu^2)\omega_0^2+\omega_0^4\right]}. 
\nonumber
\end{eqnarray}
\end{subequations}

\section{Discussion}

It is instructive to relate our approach to the existing literature on
the experimentally observed flux noise and the noise of the spin
susceptibility~\cite{Harris08,SendelbachMcDermott1}. The main features
of the experimental results are often explained with the help of the
generic model of paramagnetic
spins~\cite{DuttaHorn,McDermott,bialczak07}. The model consists of an
ensemble of non-interacting spins, each coupled to a dissipative
environment.  It is assumed that the corresponding relaxation rates
$\Gamma$ vary with the distribution function in a certain interval
between two cutoff scales $\Gamma_L$ and $\Gamma_H$:
\begin{equation}\label{p}
p(\Gamma)=\frac{1}{\ln(\frac{\Gamma_H}{\Gamma_L})} \frac{1}{\Gamma}\ .
\end{equation}
As a single spin has the Lorentzian-shaped noise spectrum, averaging
over $p(\Gamma)$ one obtains a $1/f$-noise spectrum of the whole
system within the frequency range $\Gamma_L<f<\Gamma_H$. The resulting
noise is roughly independent of temperature~\cite{McDermott} and may
be used to fit the experimental data. Weak deviations of the exponent
$\alpha$ of the measured noise
spectra~\cite{WellstoodClarke,Anton12,Drung11,Slichter12}
$S_\phi\propto 1/f^\alpha$ can be accounted for by changing the
distribution function \cite{weissman88} to $p(\Gamma)\propto
1/\Gamma^\alpha$.

The paramagnetic model does not include interactions between
spins. Indeed~\cite{Faoro12}, typical interaction scales seem to be
small, of the order $J_{\rm{typ}}\sim 50$ mK, justifying the approach
of Refs.~\onlinecite{McDermott,bialczak07} in the high-temperature
regime $T>J_{\rm{typ}}$. This is also consistent with indications that
the system of spins is in the classical high-temperature regime
characterized by a Curie
susceptibility~\cite{Harris08,SendelbachMcDermott1} and an ohmic
environment~\cite{lanting11}.

A system with a large number of non-interacting spins is a natural
generalization of our approach. In this case, instead of the single
spin-$1/2$ we need to consider the total spin of the system
\[
\hat{\bs{S}}=\sum_i^N \hat{\bs{s}}_i.
\]
The corresponding four-point correlation function
[c.f. Eq.~(\ref{SCF})] can be decomposed as follows:
\begin{eqnarray}\label{eq:CchiTotal}
&&
C_{\chi}(t_1,t_1',t_2,t_2') = - 
\av{ {\mc T}_K \hat S_z^{cl}(t_1) \hat S_z^q(t_1') \hat S_z^{cl}(t_2) \hat S_z^q(t_2')} 
\nonumber\\
&&
\nonumber\\
&&
\; \;
= - \sum_i^N \av{ {\mc T}_K \hat s_{z,i}^{cl}(t_1) \hat s_{z,i}^q(t_1') 
\hat s_{z,i}^{cl}(t_2) \hat s_{z,i}^q(t_2')} 
\nonumber\\
&&
\nonumber\\
&&
\; \;
- \sum_{i \ne j}^N \av{ {\mc T}_K \hat s_{z,i}^{cl}(t_1) \hat s_{z,i}^q(t_1') } 
\av{ {\mc T}_K  \hat s_{z,j}^{cl}(t_2) \hat s_{z,j}^q(t_2')} 
\nonumber\\
&&
\nonumber\\
&&
\; \;
- \sum_{i \ne j}^N \av{ {\mc T}_K \hat s_{z,i}^{cl}(t_1) \hat s_{z,i}^q(t_2') } 
\av{ {\mc T}_K  \hat s_{z,j}^{cl}(t_2) \hat s_{z,j}^q(t_1')}.
\end{eqnarray}
Clearly, the last two lines of Eq.~(\ref{eq:CchiTotal}) do not
contribute to Eq.~(\ref{c2}) and therefore the noise of the
susceptibility of the system of independent spins is given by the sum
of the individual noises of each spin
\begin{equation} 
X^{(2)}_{\varphi_1,\varphi_2}= \sum_i \chi^{(2)}_{\varphi_1,\varphi_2}(\Gamma_i).
\end{equation}
Averaging over the distribution (\ref{p}) one
obtains~\cite{weissman88,KochClarke,McDermott}
\begin{equation}
\label{noisemodel}
X^{(2)}_{\varphi_1,\varphi_2} = N \int^{\Gamma_H}_{\Gamma_L} d\Gamma\, p(\Gamma)\, 
\chi^{(2)}_{\varphi_1,\varphi_2}(\Gamma).
\end{equation}
Using our results (\ref{ChiEqAll}) and (\ref{ChiNeqAll}) we can now
obtain the noise of the susceptibility in the model of non-interacting
spins. In the limit, where the probing frequency $\omega_0$ is much
smaller than the slowest relaxation rate of the spins $\omega_0 \ll
\Gamma_L$ we find
\begin{eqnarray*}
X^{(2)}_{0,0} \approx  \frac{N}{4\Gamma_L} \left[\frac{2}{B_0^2}  - \frac{3}{8T^2}\right]  
{f}\left(\frac{\pi\nu}{\Delta\omega}\right) \ln^{-1}\frac{\Gamma_H}{\Gamma_L},
\end{eqnarray*} 
\begin{eqnarray*}
X^{(2)}_{\frac{\pi}{2},\frac{\pi}{2}} \approx  \frac{N}{4\Gamma_L} 
\left[\frac{2}{B_0^2}  - \frac{1}{8T^2}\right]  
{f}\left(\frac{\pi\nu}{\Delta\omega}\right) \ln^{-1}\frac{\Gamma_H}{\Gamma_L},
\end{eqnarray*} 
\begin{eqnarray}
&&
X^{(2)}_{0,\frac{\pi}{2}} \approx - \frac{N \omega_0}{4\Gamma_L^2} 
\left[\frac{3}{16 T^2}+ \frac{i\nu}{\Gamma_L}\left(\frac{4}{3B_0^2}  
- \frac{1}{6T^2}\right)\right] 
\nonumber\\ 
&& 
\nonumber\\
&&
\qquad\qquad 
\times  
{f}\left(\frac{\pi\nu}{\Delta\omega}\right) \ln^{-1}\frac{\Gamma_H}{\Gamma_L},
\end{eqnarray} 
In the opposite limit $\Gamma_L \ll \omega_0 \ll \Gamma_H$ we obtain
\begin{eqnarray*}
X^{(2)}_{0,0} \approx  \frac{\pi N}{4\omega_0} \left[\frac{1}{B_0^2}  - \frac{1}{16 T^2}\right]  
{f}\left(\frac{\pi\nu}{\Delta\omega}\right) \ln^{-1}\frac{\Gamma_H}{\Gamma_L},
\end{eqnarray*} 
\begin{eqnarray*}
X^{(2)}_{\frac{\pi}{2},\frac{\pi}{2}} \approx  \frac{\pi N}{4\omega_0} 
\left[\frac{1}{B_0^2}  - \frac{1}{16 T^2}\right]  
{f}\left(\frac{\pi\nu}{\Delta\omega}\right) \ln^{-1}\frac{\Gamma_H}{\Gamma_L},
\end{eqnarray*} 
\begin{eqnarray}
&&
X^{(2)}_{0,\frac{\pi}{2}} \approx - \frac{N}{4 \omega_0} 
\left[\frac{1}{16 T^2}+ \frac{i\pi\nu}{\omega_0}\left(\frac{1}{B_0^2}  
- \frac{1}{16T^2}\right)\right] 
\nonumber\\ 
&& 
\nonumber\\
&&
\qquad\qquad 
\times  
{f}\left(\frac{\pi\nu}{\Delta\omega}\right) \ln^{-1}\frac{\Gamma_H}{\Gamma_L}.
\end{eqnarray} 

The above results do not show the $1/\nu$ behavior observed in
Ref.~\onlinecite{SendelbachMcDermott2}. Moreover, the experiment shows
non-vanishing correlations between the fluctuations of flux and
susceptibility. As the model of independent spins remains invariant
under time reversal, such correlations are excluded in this theory. It
is therefore apparent that the model of independent spins misses the
essential physics of the real noise sources affecting SQUIDs.

Recently, J. Atalaya, J. Clarke and the two of us~\cite{Atalaya} have
performed a numerical analysis of interacting spin systems in the
presence of disorder.  It was found that the slow dynamics of the
magnetization is dominated by spontaneously forming spin clusters,
which give rise to $1/f$ noise of magnetization. We conjecture, that
the observed $1/\nu$ noise of the spin susceptibility is due to slowly
switching clusters, which affect the susceptibility of nearby spins.
The apparent time-reversal symmetry breaking could then be attributed
to the relatively short measurement times, during which some clusters
never flip their magnetization.

To conclude, in this paper we have (i) given a general definition of
noise of susceptibility and distinguished it from the second noise;
(ii) computed the noise of susceptibility of a single spin $1/2$
immersed in a dissipative environment; (iii) further developed the
powerful technique~\cite{Mao03,Shnirman03,sachdev,spencerd,spencer}
based on the Majorana-fermion representation of a spin-$1/2$
system~\cite{Martin,Tsvelik,Berezin77,shastri}; and (iv) estimated the
noise of susceptibility for the Dutta-Horn~\cite{DuttaHorn} model of
independent spins. The Dutta-Horn model appears to be insufficient to
account for all features observed in
Ref.~\onlinecite{SendelbachMcDermott2}. We conjecture that strong
spin-spin interactions, leading to cluster formation and glassy
behavior are the key ingredients needed to explain the experiment.

\begin{acknowledgments}
We wish to acknowledge discussions with A. Mirlin, Yu. Makhlin, and
J. Clarke.  We acknowledge support of the DFG under grants SCHO
287/7-1 and SH 81/2-1.
\end{acknowledgments}

\appendix*

\section{Noise of Noise}
\label{nn}

Here we briefly review the four-point correlation function
corresponding to the second spectrum or noise of noise
\cite{weissman88,kogan}. For more details the reader is referred to
the book by S.M. Kogan~\cite{kogan}.

The second spectrum was introduced in order to identify
non-Gaussian contributions in
$1/f$-spectra~\cite{kogan,SeidlerSolin,weissman88}. Let $x(t)$ be a
classical fluctuating quantity, which is the signal to be measured. In
a typical experimental protocol, the signal is bandwidth filtered and
then squared. To facilitate the comparison of the second spectrum to
the noise of susceptibility discussed in the main text, we assume
(differing from Ref.~\onlinecite{kogan}) the filter output signal of
having a form similar to Eq.~\eqref{chin}
\begin{equation}
\label{dxt}
\delta x(\tau\vert \omega_0,\Delta\omega) = 
\frac{1}{T_s} \int\limits^{\tau +\frac{T_s}{2}}_{\tau-\frac{T_s}{2}}dt\, e^{i\omega_0 t} \delta x(t).
\end{equation}
The time $T_s$ is similar to $T_\chi$ in Eq.~\eqref{chin}. This is the
time of a single measurement of the spectral density, which defines
the bandwidth $\Delta \omega = 2\pi/T_s$. The above defined $\delta x$
is related to the noise spectral density $S_x(t-t')=2 C^{(2)}(t-t')=2
\langle \delta x(t) \delta x(t') \rangle$ by \footnote{In order to
  keep the discussion as simple as possible we use just one
  exponential in Eq.~\eqref{dxt} instead of the cosine in
  Eq.~\eqref{chin} in the main text. As a consequence, we have to add
  a factor of 2 to the expressions $\lvert\delta x\rvert^2$ compared
  to the notations in Ref.~\cite{kogan}.}
\begin{eqnarray}
&&
2\langle \lvert \delta x(t\vert \omega_0,\Delta\omega)\rvert^2 \rangle 
= \int \frac{d\omega}{2\pi} 
{f}\left(\frac{\pi(\omega_0-\omega)}{\Delta\omega}\right) S_x(\omega) 
\nonumber\\
&&
\nonumber\\
&&
\qquad\qquad\qquad\qquad
\approx \frac{\Delta\omega}{2\pi}\, S_x(\omega_0).
\end{eqnarray}
The so-called second spectrum $S^{(2)}_x$ is a measure of fluctuations
of the noise power. The definition reads~\cite{kogan}
\begin{eqnarray}
&&
S^{(2)}_x(\nu\vert\omega_0,\Delta\omega) = \frac{8}{T_t}
\Big\langle \Big\lvert \int\limits^{T_t/2}_{-T_t/2}d\tau\, 
e^{i\nu \tau}  \Big(\lvert\delta x(\tau \vert \omega_0,\Delta\omega)\rvert^2 
\nonumber\\
&&
\nonumber\\
&&
\qquad\qquad\qquad\qquad
- \langle\lvert \delta x(\tau\vert \omega_0,\Delta\omega)\rvert^2 \rangle \Big)
\Big\rvert^2 \Big\rangle .
\label{S2def}
\end{eqnarray}
The time $T_t$ is the total measurement time and one can safely use
the limit $T_t\rightarrow\infty$. It is easy to show that the
following relation holds
\begin{eqnarray}
&&
\langle \lvert\delta x(\tau_1\vert \omega_0,\Delta\omega)\rvert^2 \;
\lvert\delta x(\tau_2\vert \omega_0,\Delta\omega)\rvert^2 \rangle =
\\
&&
\nonumber\\
&&
\quad\quad
= \frac{1}{T_s^4} \int\limits^{\tau_1 +\frac{T_s}{2}}_{\tau_1-\frac{T_s}{2}}dt_1dt_1' 
\int\limits^{\tau_2 +\frac{T_s}{2}}_{\tau_2-\frac{T_s}{2}}dt_2dt_2'\, 
e^{i\omega_0 ( t_1- t_1')} e^{i\omega_0 (t_2-t_2')}
\nonumber\\
&&
\nonumber\\
&&
\qquad\qquad\qquad\qquad
\times  C^{(4)}_x(t_1,t_1',t_2,t_2').
\nonumber
\end{eqnarray}
Here we introduced the classical four-point correlation function
\begin{align}
 C_{x}^{(4)}(t_1,t_1',t_2,t_2') &= \av{\delta x(t_1) \delta x(t_1') \delta x(t_2) \delta x(t_2') }.
\end{align}
This can be split into the Gaussian and the non-Gaussian parts
$C_{x}^{(4)}=C_{x}^{(4,G)}+C_{x}^{(4,NG)}$.  The Gaussian part is
obtained as
\begin{eqnarray} 
\label{C4G}
&& 
C_{x}^{(4,G)}(t_1,t_1',t_2,t_2') =  C_x^{(2)}(t_1,t_1') C_x^{(2)}(t_2,t_2') 
\\
&&
\nonumber\\
&&
\qquad\qquad\qquad\qquad\qquad
+ C_x^{(2)}(t_1,t_2) C_x^{(2)}(t_1',t_2')  
\nonumber\\  
&&
\nonumber\\
&&
\qquad\qquad\qquad\qquad\qquad
+ C_x^{(2)}(t_1,t_2')\ C_x^{(2)}(t_1',t_2).
\nonumber
\end{eqnarray}
One can now use Eq.~\eqref{C4G} to find the Gaussian contribution to
the second spectrum. One finds that the first term on the right-hand
side of Eq.~\eqref{C4G} cancels the average $\langle\lvert\delta
x\rvert^2 \rangle$ in Eq.~\eqref{S2def}. The remaining two terms
contribute yield the Gaussian contribution to the second noise
spectrum
\begin{eqnarray}
\label{GaussNoise}
&&
S^{(2)}_{x,G}(\nu\vert\omega_0,\Delta\omega) = 8 \int \frac{d\Omega}{2\pi}\, 
C_x^{(2)}(\Omega) C_x^{(2)}(\Omega+\nu)
\\ 
&&
\nonumber\\
&&
\qquad\qquad\qquad\quad 
\times 
{f}\left(\frac{\pi(\omega_0-\Omega)}{\Delta\omega}\right) 
{f}\left(\frac{\pi(\omega_0-\nu-\Omega)}{\Delta\omega}\right).
\nonumber 
\end{eqnarray}
In contrast, the noise of susceptibility does not contain any
contribution of the Gaussian part of the 4-point correlation function
\eqref{Gdecomp}.

In the experimentally relevant regime $\nu\ll\Delta\omega\ll\omega_0$
we can approximate Eq.~(\ref{GaussNoise}) as
\begin{equation}
S^{(2)}_{x,G}(\nu\vert\omega_0,\Delta\omega) \approx
\frac{8\Delta\omega}{3\pi}\, \left(C_x^{(2)}(\omega_0)\right)^2.
\label{GaussNoiseAppr}
\end{equation}
This spectrum is "white" as a function of $\nu$.

Now, in the system of $N$ non-interacting spins the noise spectrum
scales linearly with the number of spins $S_s(\omega)\propto N$, while
the Gaussian contribution to the second noise scales as
$S^{(2)}_G(\nu|\omega,\Delta\omega)\propto N^2$, see
Eq.~\eqref{C4G}. In contrast, the non-Gaussian contribution is
linear in $N$, similarly to Eq.~(\ref{noisemodel}). Therefore, the
Gaussian part of the second noise always dominates making it difficult
to extract information about non-Gaussian fluctuations from the second
noise measurements.

The above conclusion is not necessarily general. In context of Ising
spin glasses, it has been shown \cite{Nguyen01,Chen10} that the
infinite-range interaction may result in a $1/f$ second noise spectrum
in the thermodynamic limit $N\rightarrow\infty$.

\bibliography{mybibfile}

\end{document}